\documentclass[%
 aip,
 rsi,
 amsmath,amssymb,
 reprint,%
]{revtex4-1}

\usepackage{graphicx}
\usepackage{dcolumn}
\usepackage{bm}

\usepackage[utf8]{inputenc}
\usepackage[T1]{fontenc}
\usepackage{mathptmx}
\usepackage[lofdepth,lotdepth]{subfig}
\usepackage{listings}
\usepackage{hyperref}
\begin{document}



\title{Acoustic Waves in Granular Packings at Low Confinement Pressure} 



\author{Karsten Tell}
\email[]{karsten.tell@dlr.de}
\affiliation{Institut f\"ur Materialphysik im Weltraum,
  Deutsches Zentrum f\"ur Luft- und Raumfahrt (DLR), 51170 K\"oln, Germany}

\author{Christoph Drei{\ss}igacker}
\affiliation{Institut f\"ur Materialphysik im Weltraum,
  Deutsches Zentrum f\"ur Luft- und Raumfahrt (DLR), 51170 K\"oln, Germany}

\author{Alberto Chiengue Tchapnda}
\affiliation{Institut f\"ur Materialphysik im Weltraum,
  Deutsches Zentrum f\"ur Luft- und Raumfahrt (DLR), 51170 K\"oln, Germany}

\author{Peidong Yu}
\affiliation{Institut f\"ur Theoretische Physik, Universit\"at zu K\"oln, D-50937 Cologne, Germany}

\author{Matthias Sperl}
\affiliation{Institut f\"ur Materialphysik im Weltraum,
  Deutsches Zentrum f\"ur Luft- und Raumfahrt (DLR), 51170 K\"oln, Germany}
\affiliation{Institut f\"ur Theoretische Physik, Universit\"at zu K\"oln, D-50937 Cologne, Germany}


\date{\today}

\begin{abstract}
Elastic properties of a granular packing show nonlinear behavior determined by its discrete structure and nonlinear inter-grain force laws. Acoustic waves show a transition from constant, pressure-dependent sound speed to a shock-wave like behavior with amplitude-determined propagation speed. This becomes increasingly visible at low static confinement pressure as the transient regime shifts to lower wave amplitudes for lower static pressure. In microgravity, confinement pressure can be orders of magnitude lower than in a ground based experiment. Also, the absence of hydrostatic gradients allows for much more homogeneous and isotropic pressure distribution. We present a novel apparatus for acoustic wave transmission measurements at such low packing pressures. A pressure control loop is implemented by a microcontroller that monitors static force sensor readings and adjusts the position of a movable wall with a linear motor until the desired pressure is reached. Measurements of acoustic waves are possible using accelerometers embedded in the granular packing as well as piezos. For excitation we use a voice coil-driven wall, with a large variety of signal shapes, frequencies and amplitudes. This enables experiments both in the linear and strongly nonlinear regime.
\end{abstract}


\maketitle 



%
%

%

\section{Introduction}
\label{sec:intro}
Defining an unambiguous speed of sound in granular packings is nontrivial. First, let us examine the low amplitude case, where the excitation signal in terms of dynamic pressure is small compared to the static confinement pressure of the packing. Liu and Nagel\cite{Liu_Nagel} found a huge discrepancy of measured time-of-flight speed ($\approx$ 280 m/s) and group velocity ($\approx$ 60 m/s) calculated from the phase-difference as a function of frequency. Jia et al.\cite{Jia1} explained the difference by taking into account the finite sensor size and its ratio to particle size. A sensor with large area in contact with many particles averages over sound transmitted through many different paths, resulting in a self-averaging or coherent signal. Its signal shape is reproducible between different realizations of packings at the same pressure. The coherent signal is well described by Effective Medium Theory\cite{Walton} which predicts a compressional wave speed $v_p\propto \Phi^{-1/6} Z^{1/3} P^{1/6}$ which is consistent with measurements of time-of-flight speed. Here $\Phi$ is the volume fraction, $Z$ the average coordination number and $P$ the confinement pressure of the packing and the wavelength $\lambda$ is much larger then the particle diameter $d$. This behavior was found in measurements in soils of many types and is well known in soil mechanics, geophysics and engineering\cite{Digby,Goddard}.

On the other hand, if the sensor is in contact with only one or few particles, the measured signals contain a strong contribution of highly irregular shape that is not reproducible even between packings prepared with the same protocol at the same static pressure. This is referred to as incoherent signal\cite{Jia1}. For an initial short excitation pulse the received signal in general contains a coherent part resembling the attenuated, broadened pulse, superimposed by or followed by an incoherent part in the form of a high-frequency tail or coda of much longer duration and higher bandwidth than the original excitation signal\cite{Coste, Jia2}. This coda signal is considered the sum of sound propagating through all possible paths through the granular medium between emitter and receiver. Such paths cans be provided by the network of force-carrying links between neighboring particles known as force-chains\cite{travers,radjai,jaeger}, which have been actively studied in the literature. As the length and pre-compression of these force-chains varies, waves traveling along them acquire different relative phases. The resulting interference or acoustic speckle pattern is highly sensitive to changes in the force-distribution of the packing. Evidence for sound transmission predominantly through force-chains was given by Daniels et al.\cite{Daniels1} who measured acoustic waves in a layer of stress-birefringent disks. Changes in the force-chain network can thus be seen as a drop in the cross-correlation of subsequently measured coda signals\cite{Tournat1,Jia3}. For increasing excitation frequency the incoherent contribution to the received signal is increasing while the coherent contribution is decreasing as a result of multiple scattering, until for $\lambda \ll d$ diffusive wave propagation is found\cite{Jia2,Trujillo}.

For propagation of high amplitude pulses we have to examine the role of nonlinear contact forces i.e. deviations from Hooke's law due to $F\propto\delta^\alpha$ with $\alpha\neq1$, opening and closing of contacts or hysteresis due to frictional sliding and viscoelastic damping. For zero static pre-compression force in a chain of beads, a solitonic propagating wave results as a solution of a nonlinear wave equation as shown by Nesterenko\cite{Nesterenko}. The spatial pulse-width of such a soliton is five particle diameters and the group speed is given as a power-law function of the amplitude. Such behavior was found experimentally in chains of beads\cite{Daraio_1dchain}. In 2D systems with low pre-compression similar behavior arises, where a step-like excitation leads to a propagating shock-wave with a soliton-like wavefront followed by an oscillating tail as shown in simulations by Gomez and Vitelli et al.\cite{Gomez}. The presence of disorder leads to decay of the wavefront by redistribution of energy from the front to the tail as a function of propagated distance. Such shock-waves have been found experimentally by van den Wildenberg et al.\cite{Wildenberg} in a packing of glass beads at different confinement pressures. As the ratio of wave amplitude to static pressure was varied by several orders of magnitude, a smooth transition from constant to amplitude-dependent wavefront speed was found, in accordance with the Hertzian contact force law. Similar experiments confirmed these observations\cite{Santibanez}. These findings are all examples of strongly nonlinear behavior of the granular packing itself.

Attenuation of granular waves also strongly depends on the wave amplitude. For low-amplitude waves an exponential decay of amplitude over propagation distance is found. However, for high amplitude shock-like waves increasing attenuation as a function of amplitude is found\cite{Wildenberg} which can be fitted by a power-law relation between sound pressures measures at different propagation distances.

While numerous experiments have been conducted in simplified 1D or 2D model systems of granular packings at low pressure close to unjamming, such as linear chains\cite{Daraio_1dchain} or horizontal layers of beads\cite{Coste}, similar studies of 3D systems are mostly missing. Notable exceptions include a study of time-of-flight speeds of pressure- and shear waves under microgravity in a drop tower campaign\cite{Zeng} without any measurement or adjustment of confinement pressure. More recently, an apparatus for a broad variety of granular physics experiments has been proposed\cite{aumaitre2018} to potentially be equipped with a cell dedicated to sound measurements at adjustable confinement pressure. In microgravity, packing pressures much smaller than the hydrostatic pressure can be achieved by imposing a well-defined external confinement force. Here the lower bound is given by the accuracy of the pressure control loop implemented in the experimental apparatus. Such an apparatus must be capable of packing preparation according to a protocol that reliably creates a stable packing. For measurements of shock-wave-like behavior we need an excitation system of sufficient strength, while for measurements of harmonic waves sufficiently high bandwidth and low signal distortion is required. For meaningful measurements, sufficient time in microgravity is required. While drop towers, such as the ZARM facility, and parabolic flights with modified aircraft, such as provided by Novespace, offer up to 9 s in the former case and up to 22 s in the latter case, this is barely enough for repeated sound measurements when 5 to 10 s of time is taken into account for the packing preparation protocol. The Mapheus sounding rocket of DLR enables us to use 6 minutes of microgravity, which is sufficient for measurements at several confinement pressure settings. The residual acceleration shall be kept below $10^{-4}$ g as was previously achieved during flights lasting more than three minutes\cite{mapheus}. Under these conditions measurements close to unjamming and close to the sonic vacuum shall be possible. In this work, we present a novel apparatus, implemented as a Mapheus module, to conduct such measurements.
\section{Experimental Setup Overview}
\label{sec:overview}
\begin{figure}
\includegraphics[width=\hsize]{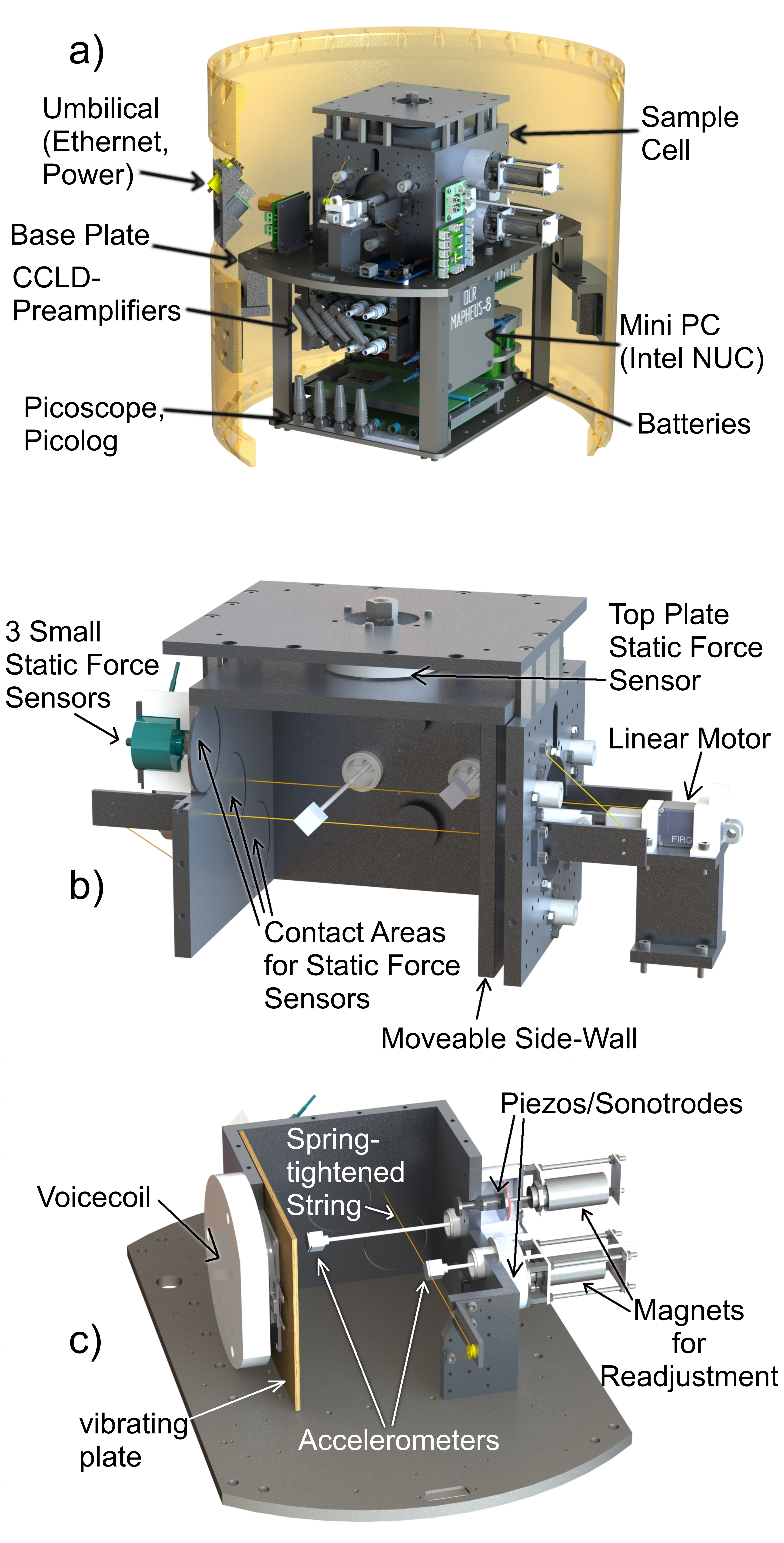}

\caption{Granular Sound module: (a) complete apparatus inside the module container for the Mapheus sounding rocket. The upper part contains the sample cell, sensors, excitation system and pressure control system. The lower part contains pre-amplifiers, data-logger, oscilloscope and mini-PC as well as and batteries. (b) pressure control system with three force sensors embedded in a side-wall, one force sensor on the top and a linear-motor driven movable side-wall to compress the packing up to the desired confinement pressure. (c) sound measurement system with voice-coil-driven vibrating wall, two accelerometers inside the packing at 33 and 88 mm distance from the vibrating wall and two piezoelectric sensors mounted at the side-wall opposite to the vibrating wall with different areas in contact with the packing.}
\label{fig:cad}
\end{figure}

Our apparatus is shown in Fig. \ref{fig:cad}. It consists of two modular parts mounted on top of each other. The upper part has a base plate that is mounted inside a single unit module container for the Mapheus rocket. On top of the base plate there is the box-shaped, aluminium-made sample cell with internal dimensions 12 cm x 12 cm x 13 cm, filled with spherical glass beads. We use a bidisperse mixture of 4 and 3 mm beads at mass ratio 1:1, unless mentioned otherwise in the following sections. The inner surface of all cell walls is padded with soft foam to decouple the sample from vibrations of the structure and to minimize reflections at the boundaries. The foam stiffness is small compared to the effective stiffness of the granular packing and to the stiffness of the beads constituting the packing. Therefore the foam is flexible enough to compensate small local fluctuations in the bead indentation, effectively keeping the packing boundaries at constant pressure rather than constant position. For acoustic excitation there is a voice-coil (Visaton EX 80 S) driving a plate made of glass-fiber reinforced plastic. Apart from a 1 mm gap on all sides, the plate covers the cross sectional area of the cell and is in direct contact with the beads. Orthogonal to the plate there is a movable side-wall, driven by a linear-motor (Actuonix L12) to compactify the packing to provide an adjustable confinement pressure (see next section). The maximum displacement of this wall is 11 mm, leading to an increase in volume fraction by 9\%.

On the opposite side-wall there are three static force sensors (Burster 8432-5005 strain-gauge sensors) arranged in a diagonal pattern. Each sensor is in contact with the beads via a aluminium disk of 43 mm diameter padded with foam. The top plate is mounted to a static force sensor (Burster 8432-5050) which is mounted on the cell. The plate covers the cross-sectional area of the cell apart from a 1 mm gap and is also covered with foam on the inside. Inside the sample cell there are two accelerometers (Brüel\&Kjær 4508-B) located at 32 and 88 mm distance from the excitation plate at half the height of the cell. They are held in place by strings crossing the sample cell which are kept under constant low tension by metal springs on the outside of the cell. The tension is adjusted low enough that for the largest measured sensor displacement due to elastic wave transmission, $\approx$ 1 $\mu$m, the restoring force, $\approx$ 80 $\mu$N, will be much lower than the force acting on the sensor due to the pressure wave, $\approx$ 1 mN. To verify this, we transversally displaced the string by 5 mm while measuring a restoring force of 0.4 N. The string diameter $\ll$ 1 mm is much smaller than the bead diameter 3 - 4 mm, reducing any effect of the string on the local packing structure as well its scattering cross section as much as possible. An additional accelerometer of the same type can be mounted directly on the excitation plate for test measurements to record its motion. Alternatively, two piezo-electric sensors are used which are mounted at the cell wall opposite to the voice-coil. They consist of a piezo-ceramic disk of 2 nF capacitance, fundamental resonances at 1.8 MHz (thickness mode) and 105 kHz (radial mode) attached to a machined aluminium disk of 25 and 12.5 mm diameter. To decouple them from the structure the piezo-disks are suspended between layers of foam inside a metal housing while aluminium disks are in direct contact with the beads.

The static force sensors are connected to a preamplifier (Burster 9236) which is connected to a an analog-digital converter(ADC) and data-logger (Pico Technology ADC-24) for high-precision readout and to the ADC of a microcontroller(Arduino Uno) for fast readout for the pressure control loop (see next section). It controls the linear-motor via a dedicated motor driver. Optocouplers are used to convert the 28 V signals provided by the Mapheus service module to 5 V signals which are then read out by the microcontroller. The signals are used to indicate a switch from external to internal power as well as liftoff of the rocket and start of the experiment in microgravity.

The accelerometers and piezos are connected to constant current (CCLD)-preamplifiers (Brüel\&Kjær 1704-A-002). For the piezos no bias current is used. The signals are read out by a digital oscilloscope(Picoscope 5442B). All sensor data is saved on a solid state disk of a mini-PC (Intel NUC i5). The latter runs custom written software to run all planned measurements during an experimental campaign. It reads status messages and pressure readings from the microcontroller and sends commands for pressure adjustment via serial connection over USB. The CCLD-preamplifiers, oscilloscope, data-logger and mini-PC as well as onboard power supply are contained in the lower module part. The power supply is implemented by a series of LiFePo batteries with a total nominal voltage of 24 V. They are charged from a ground-based 28 V supply provided by the umbilical. There is a diode-based circuit for uninterrupted power supply that switches from ground-based power to battery power at liftoff while keeping the experiment running. This switching process is triggered via signal from the service module which switches a relay in our module. Several DC-DC voltage converters (Traco) with radiative cooling are used to provide adequate supply voltages for all devices.

An Ethernet connection is provided by the umbilical to enable remote control and monitoring of all devices until liftoff. During the countdown, a series of test measurements is conducted automatically to verify the proper function of all components. Once these tests are completed, the software runs continuous checks of all devices and monitors the service module signals while waiting for launch. Any malfunctioning devices such as the microcontroller, the data-logger or the oscilloscope are automatically reinitialized if necessary during the countdown or the flight. Unless extraordinary errors occur, intervention by a human operator is not required. However, extensive diagnostic and debugging features are available via SSH connection. If necessary, the software and microcontroller firmware can be updated within seconds without interrupting the countdown. Once the microgravity phase has started, as indicated via signal from the service module, the software runs all planned measurements according to a series of setting files. Each file specifies the desired confinement pressure, excitation signal and device settings for the oscilloscope. All sensor readings as well as the oscilloscope waveforms are saved along with UNIX timestamps with millisecond precision according to the time of conversion or the trigger point. This enables us to assign pressure readings precisely to each waveform or within a waveform. After six minutes all measurements are stopped, the data is compressed and the NUC is shutdown to prepare for reentry. Apart from the data, extensive log entries of all software components are saved to document any malfunctions that may arise during the campaign to simplify troubleshooting if necessary.
\section{Packing Preparation}
\label{pack_prep}
Before we can conduct sound measurements, preparation of a packing at known, specified static pressure that remains stable for the duration of the measurement is required. For this purpose the sample cell has a movable side-wall, driven by a linear-motor to adjust its position. It can compress the packing up to a pressure of 2.5 kPa. Readings from static force sensors embedded in the side-walls are taken, averaged and fed to a control-loop implemented on an Arduino Uno microcontroller, which then directs the linear-motor to drive the wall back and forth in sub-millimeter steps until the specified pressure setting is reached.

\begin{figure}
\includegraphics[width=\hsize]{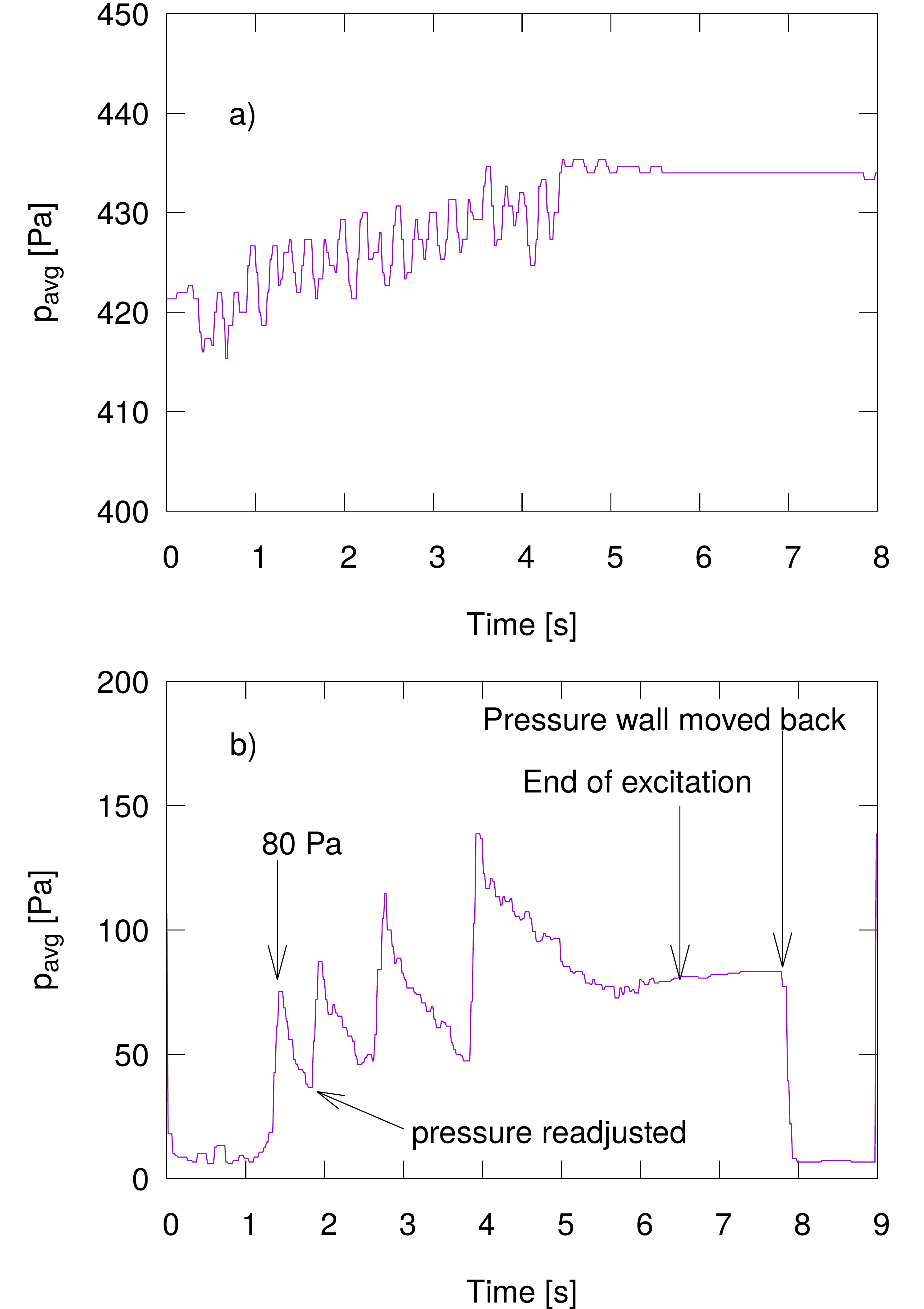}
\caption{Evolution of packing pressure as measured by force sensors in a side-wall. (a): On ground, at initial pressure 420 Pa, it increases by 15 Pa during repeated acoustic excitation. (b) In microgravity, after the initial pressure of 80 Pa is reached at 2 s, the pressure decreases during excitation of the same strength as in (a) until the lower threshold of the pressure control loop is reached. Then the pressure is readjusted and the measurement is repeated three times. }
\label{fig:zarm_press_ground_vs_mug}
\end{figure}

Our measurements on ground and in microgravity show that, while the desired pressure setting is reliably reached within 10 - 30 iterations within one second, the pressure does not remain stable under excitation by vibrations and sound transmission. Whether the pressure decreases or increases depends on the excitation strength, the packing preparation protocol and on the initial pressure just after packing preparation. This behavior is known in the literature\cite{Jia4}. For repeated acoustic excitation by short pulses of moderately high amplitude within our accessible range we observed a monotonous increase of pressure on ground but monotonous decrease in microgravity as shown in Fig. \ref{fig:zarm_press_ground_vs_mug}. In the latter case, the pressure drops without bound until the packing loses mechanical rigidity entirely. When this happens, the pressure control loop detects a drop below a predefined threshold and readjusts the position of the movable wall again. On the other hand, for excitation at the highest accessible strength we observed a decrease in pressure at all initial pressure settings. At the lowest amplitudes no change in packing pressure was observed even after many repetitions.

After our initial tests of packing preparation protocols in the ZARM drop tower and on the NOVESPACE parabolic flight plane we developed the following protocol to achieve a packing configuration that remains at stable pressure during sound transmission even at high amplitudes:
First, the wall is moved outwards to loosen the packing until, in microgravity, the packing unjams and packing pressure becomes zero. This state is used as reference point to remove offsets from the force sensors. Then, while using the offset-corrected readings from this point on, the wall is moved inwards until the specified pressure plus an overshoot of 20 \% is reached. Finally, a series of strong pulses is transmitted by the voice-coil, to vibrate the packing for at least 10 seconds. This time constraint makes the protocol unsuitable for use in drop tower campaigns, but still suitable for parabolic flight or sounding rocket campaigns. The pulse amplitude, measured in the center of the packing, is 1 $m/s^2$ or 500 Pa, which is sufficient to trigger rearrangements in the particle positions and the force distribution, visible as changes in the static pressure as measured in all force sensors embedded in the sample cell walls. Simultaneously, continuous readjustment of the wall position is taking place, in an attempt to keep the average pressure at the specified value. The readjustment steps, calculated by the microcontroller to be proportional to the pressure error, become increasingly smaller. During this process the overshoot is lowered exponentially with time. Continuous readjustment ceases when the error drops below a specified threshold of 5 Pa or when it reaches a timeout at 180 iterations. From this point on, packing pressure is being monitored but not readjusted, unless it deviates from the specified value by more than 100 Pa or it drops below the absolute minimum threshold of 20 Pa. This condition is checked during sound measurements before each waveform.

We found that for repeated preparation of packings at the same specified pressure, the packing fraction stays within 0.15 \%. This was measured by alternating eight times between 500 and 800 Pa.

\begin{figure}
\resizebox{\hsize}{!}{\includegraphics{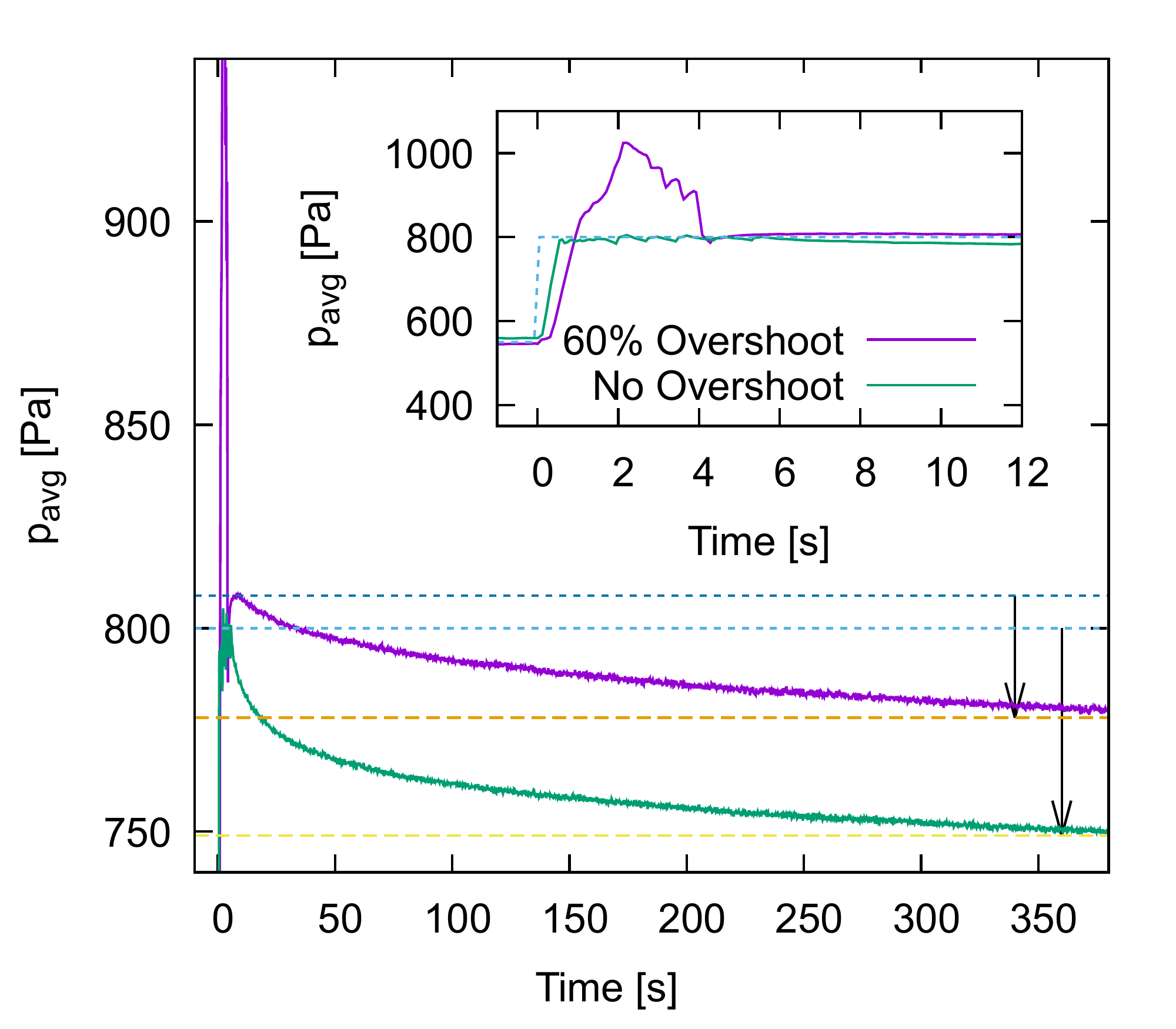}}
\caption{Confinement pressure vs. time after packing preparation with two different protocols: a side wall position is adjusted continuously with a linear-motor while the packing is vibrated by a voice-coil until the specified pressure is reached and remains stable within a specified threshold. Within few minutes a continuous pressure drop is observed. When a pressure overshoot is used (violet, upper curve) the drop is much smaller than in the case of no overshoot (green, lower curve). The inset shows both protocols in detail.}
\label{fig:pack_prep_overshoot}
\end{figure}

Following the readjustment, a slow increase or decrease towards the previous pressure value is observed. By carefully choosing an appropriately large overshoot, this long-time evolution can be largely suppressed as shown in Fig. \ref{fig:pack_prep_overshoot}. If the desired pressure setting is lower than the previous setting, an undershoot is used. While this method reliably succeeds in preparing a packing of stable pressure on the timescale of microgravity experiments - one or few minutes - an even slower pressure evolution over the course of hours and days was still observed during test-runs in the laboratory.

\section{Acoustic Excitation}
\label{sec:excitation}
A voice-coil-driven vibrating wall is used to introduce elastic waves into the packing. Apart from a gap of few millimeters to all sides, the wall covers the cross-section of the sample cell. It is made of 2 mm thick glass-fiber reinforced plastic (GFRP), which is chosen to be rigid and lightweight to enable excitation of plane waves in a large frequency range. Experimentally we find the lowest resonance at 20 kHz in agreement with the lowest Lamb mode given by $f_0=c_{GFRP}/(2 L)$ with length of the wall $L=$12 cm and material sound speed $c_{GFRP}=$5 km/s. For lower frequencies all points of the vibrating wall are in phase within $\pm \pi/2$ or lower so plane waves are generated in good approximation.

The voice-coil (Visaton Ex 80) is rigidly mounted on the side opposite to the vibrating wall. It is driven by a bipolar power-amplifier into which a signal is fed that is generated by the built-in arbitrary waveform generator (AWG) of the oscilloscope (Picoscope 5442B) used for generating and recording signals. We use software-generated waveforms such as pulses, Gaussian tone bursts, sinusoidal or chirp signals that are calculated on the NUC computer and loaded into the AWG buffer as needed during the measurement campaign. The electrical excitation signal is used to trigger the oscilloscope when a series of measurements with accurately reproducible trigger points is needed, e.g. for time-of-flight (TOF) measurements of short pulses.

As the voice-coil contains a volume of trapped air, which affects its impulse response depending on environmental pressure, we drilled millimeter-sized holes to enable the air to escape under vacuum. As a result we found a consistent impulse response during measurements in a vacuum chamber where we mounted an accelerometer rigidly on the vibrating wall.

To enhance the effectively usable bandwidth and to get a well-defined impulse-response of our excitation system we developed a software-based inverse-convolution filtering method as described in appendix \ref{sec:pulse_shaping}. By applying it to a waveform before loading it into the AWG buffer we get an improved excitation signal that resembles the desired original waveform with high accuracy up to 40 kHz of bandwidth, which we verified by measurements with an accelerometer on the vibrating wall.
\section{Time of Flight}
\label{sec:time_of_flight}
Simulations of 2D frictionless granular packings with disorder\cite{Gomez} show that the speed of shock fronts depends on the shock amplitude according to Eq. (\ref{eqn:vshocksim}), where $\delta_S$ is the indentation at the shock front, $\delta_0$ is the static indentation, $\alpha$ is the exponent of the interparticle contact potential $U \propto \delta^\alpha$ and $c$ is the linear sound speed.

\begin{align}
v_{Shock}=c \sqrt{\frac 1 {\alpha-1} \frac{(\delta_S/\delta_0)^{\alpha-1}-1}{(\delta_S/\delta_0)-1}}
\label{eqn:vshocksim}
\end{align}

For a Hertzian contact model, applicable to a packing of frictionless spheres, $\alpha=5/2$ and $v_{Shock}\propto \delta_S^{1/6}$ for $\delta_0 / \delta_S \to 0$. In this limit, called the sonic vacuum, no linear regime exists and only nonlinear excitations are possible. The transition from the weakly to the strongly nonlinear regime can be represented by the empirical relation (\ref{eqn:vshockfit}) for the propagation speed that was fitted to experimental data obtained in ground-based experiments\cite{Wildenberg}. In these experiments a lower bound for the confinement pressure or static indentation is given by hydrostatic pressure. However, in microgravity arbitrarily low pressure is accessible, limited only by technical capabilities of the experimental apparatus.

\begin{align}
v_{Fit}=c \left( 1+\frac {P_{m}} {P_{i}}\right)^{1/ 6}
\label{eqn:vshockfit}
\end{align}

\begin{figure}
\resizebox{\hsize}{!}{\includegraphics{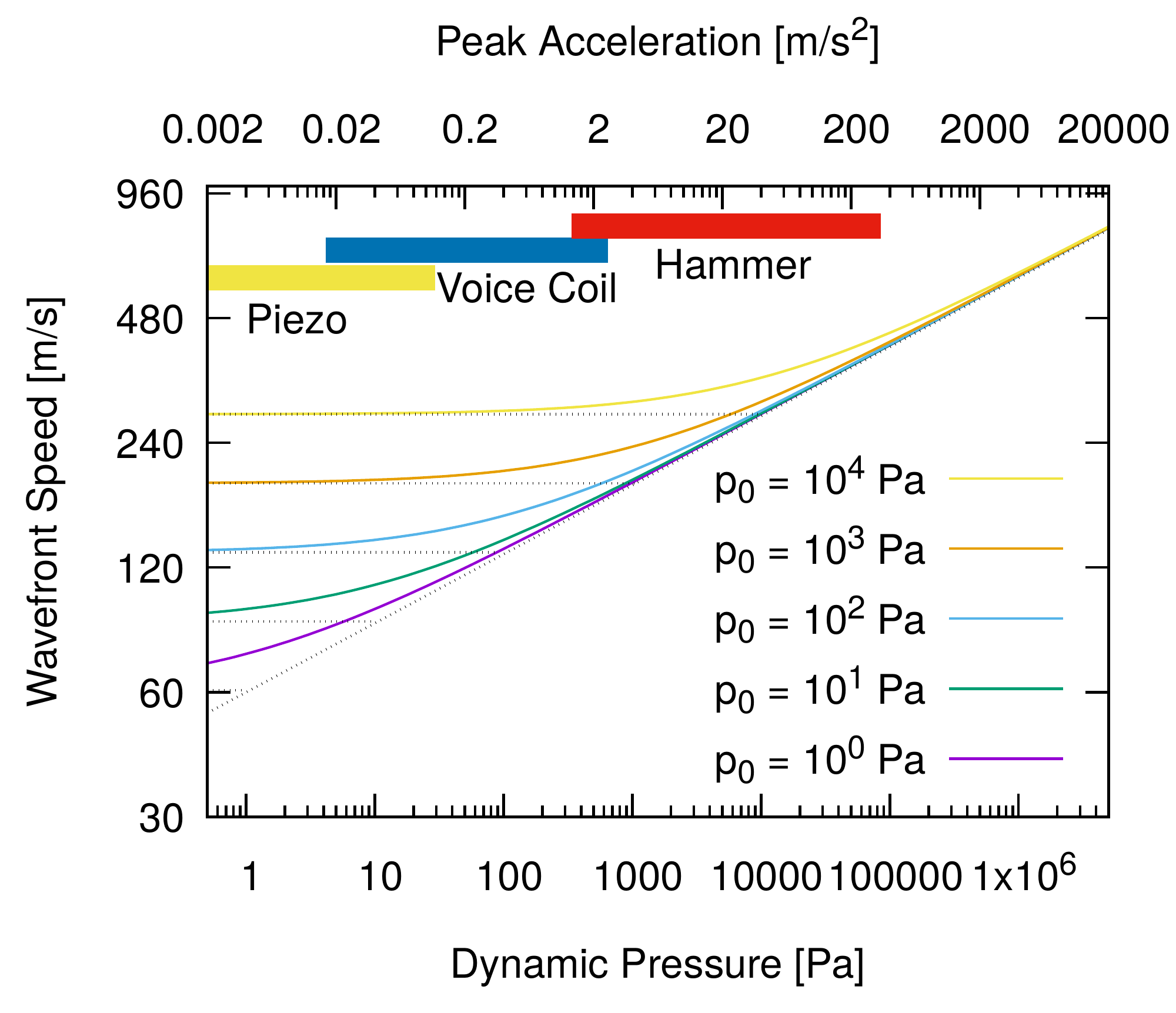}}
\caption{Predicted wave front speed vs signal strength according to Hertz-Mindlin contact force model. In our experiment the accessible pressure range is $\propto$10 Pa - 1 kPa. The amplitude range of our voice-coil (blue line) closes the gap between the range of piezo-based (yellow line) and hammer-based (red line) excitation used in previous attempts.}
\label{fig:range}
\end{figure}

The group velocity of short pressure pulses is determined by their time of flight. In a typical ground-based measurement the propagating pulses resemble Gaussians of FWHM 200 $\mu$s which corresponds to 4 cm of spatial width or 11 particle diameters if the group velocity is 200 m/s. The peak amplitude is varied by a factor of 128. The signals are detected by accelerometers embedded in the packing at 32 and 88 mm distance from the excitation wall and by two piezo transducers at 120 mm. The piezos are mounted on the sample cell wall opposite to the excitation wall. They are mechanically decoupled from the wall by soft plastic foam to reduce signal contributions from the structure. Using the geometric mean $a_m=\sqrt{a_{close}\cdot a_{far}}$ of the peak accelerations we can quantify the maximum transmitted pulse strength as 8 $m/s^2$. Using a dynamic force sensor at the same distance from the excitation wall as each accelerometer, we find a peak pressure proportional to the peak acceleration such that the equivalent geometric mean pressure satisfies $p_m=245 Pa/\frac{m}{s^2} \cdot a_m$. This relationship is found linear across the entire amplitude range. According to Eq. (\ref{eqn:vshockfit}) we should be able to measure strongly nonlinear behavior at low pressure achievable in a microgravity experiment as shown in Fig. (\ref{fig:range}).

\begin{figure}
\includegraphics[width=\hsize]{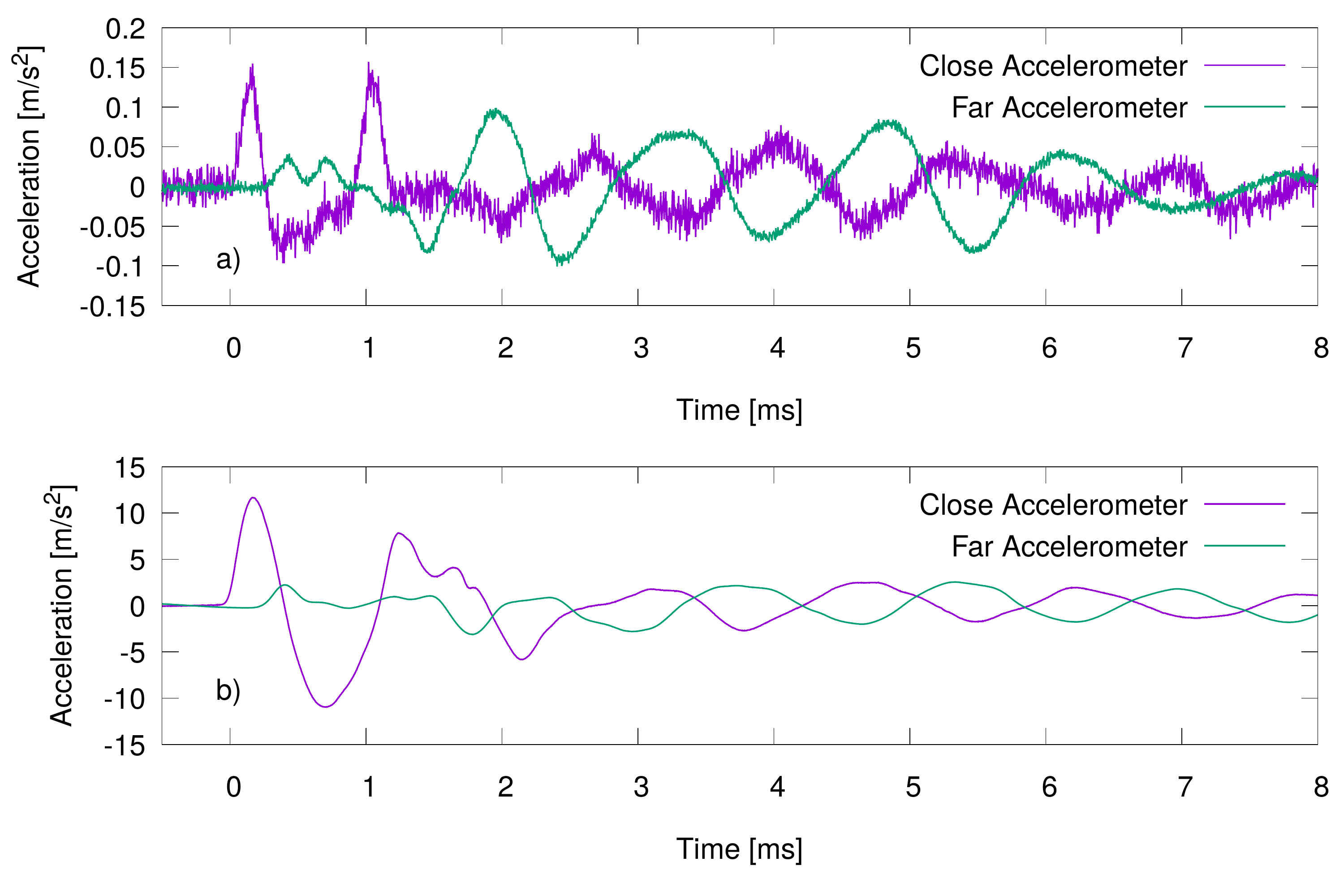}
\caption {Received signals as measured by accelerometers at 33 and 88 mm distance from the vibrating wall after halfsine excitation at (a) low and (b) high amplitude. The leading pulse is used to determine the wavefront speed.}
\label{fig:linear_and_nonlinear_signals}
\end{figure}

\begin{figure}
\includegraphics[width=\hsize]{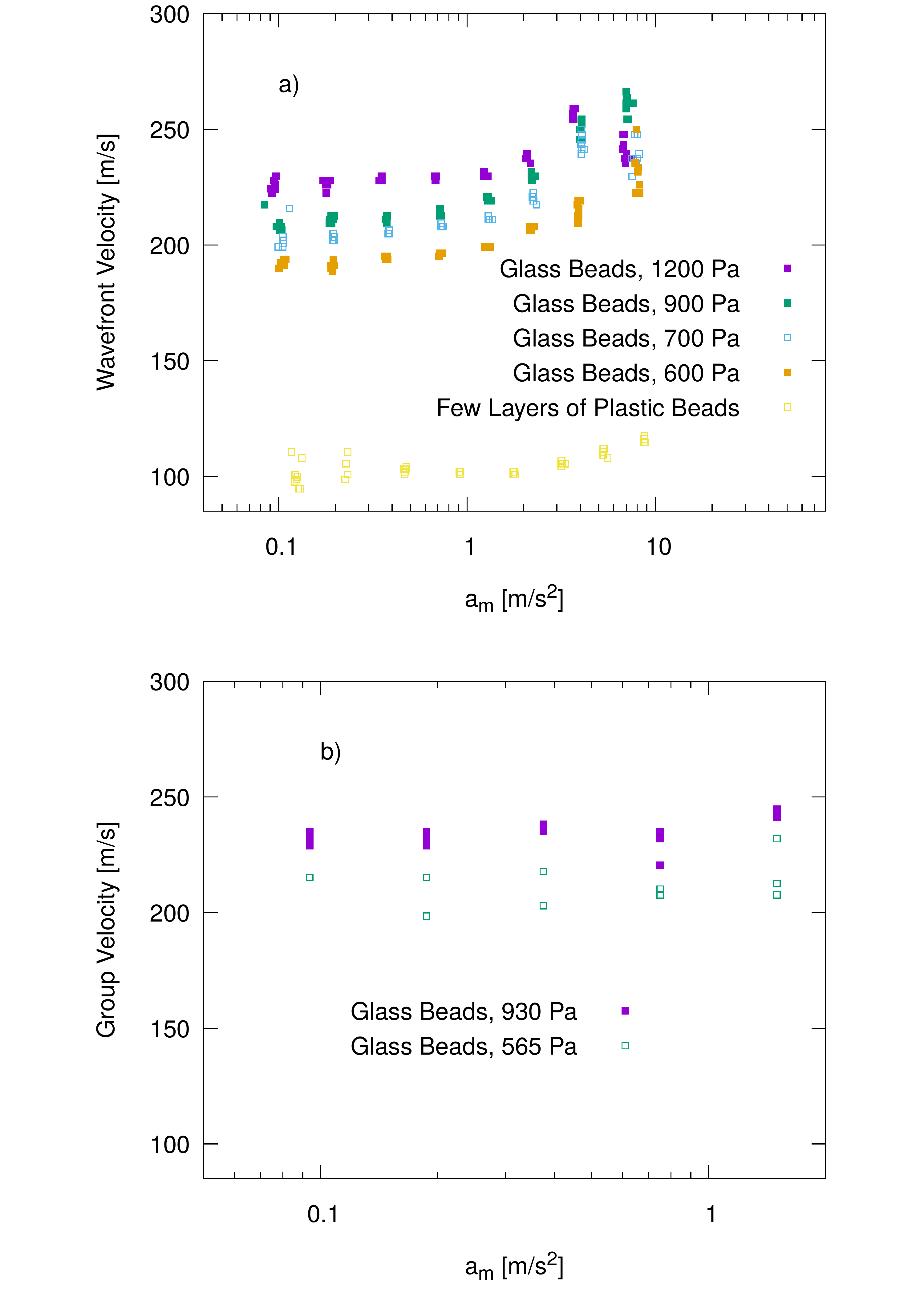}
\caption{Time-of-flight speed against amplitude (geometric mean of maximum received acceleration signals) determined for (a) short pulse signals and (b) Gaussian tone burst signals with 4 cycles of 4 kHz center frequency for different static pressure as determined by force-sensors sensors in a side-wall. In (a) two pressure settings for packings of bidisperse glass beads of 3 and 4 mm diameter are shown as well as the result for 5mm plastic beads where the cell was only partially filled. In (b) the same glass beads as in (a) were used.}
\label{fig:halfsine_and_toneburst}
\end{figure}

From the measured signals as shown in Fig. (\ref{fig:linear_and_nonlinear_signals}) we determine the time of arrival at each sensor at the pulse maximum. Compared to other methods such as using the first arrival determined by a threshold or using the first zero-crossing, this gives the most consistent results for the propagation speed, independent of emitter-receiver distance as was found by Jia et al.\cite{Jia5}. Fig. \ref{fig:halfsine_and_toneburst}a) shows the result of such analysis for short pulses propagating in packings of glass beads and plastic beads. Alternatively, the highest peak in the cross-correlation of both accelerometer signals is used. We found that both methods give consistent results that are numerically different by up to 30 \% but are following the same trend when pulse amplitude and static packing pressure are varied. For signals of more complex structure than a short pulse, like tone bursts containing several oscillation cycles, we use a FFT-based deconvolution method similar to what Jia et al used previously\cite{Jia5}. One important difference is that we use the signal measured by the close accelerometer as reference signal, as opposed to the signal measured directly on the emitting transducer as done by Jia et al. The deconvolution cancels out the sensor response as long as both sensors are identical. Both sensors are embedded deep within the packing, thus they are measuring a traveling wave. Therefore our deconvolved signal resembles the impulse response of the granular medium between the far and the close sensor while avoiding contributions from a transition between reactive and radiative regime. Fig. \ref{fig:halfsine_and_toneburst}b) shows the resulting group speed for Gaussian tone bursts propagating through packings of glass beads. It must be noted that the results are found sensitive to the choice of the region of interest for the analysis. We use a region that includes the initial peak up to the first zero-crossing but that excludes the much longer fluctuating tail following each pulse. The signal shape of this tail is found highly irregular and not reproducible for individual packings prepared at the same pressure.

A comparison of our ground based measurements as shown in Fig. \ref{fig:halfsine_and_toneburst}a) to the Hertzian prediction as shown in Fig. \ref{fig:range} shows that while we can already see the onset of increasing wavefront speed at the highest amplitude settings corresponding to the strongly nonlinear regime, the wave speed remains strongly dependent on the confinement pressure as is characteristic of the weakly nonlinear regime. The pressure dependence we measured appears to more closely resemble a $v_p\propto p_0^{1/4}$ power law instead of $v_p\propto p_0^{1/6}$. While the latter is found in jammed packings at high confinement pressure where any increase in pressure results in affine deformations that leave the coordination number $Z$ constant, this may no longer be valid for sufficiently low pressure. We have to stress that our packings prepared on ground are subject to strong pressure gradients. Thus, a meaningful investigation of the low pressure behavior needs to take place under microgravity where much more isotropic and homogenous packings could be prepared. In microgravity, where $p_0$ could be decreased by two orders of magnitude, we expect to be able to approach the sonic vacuum, characterized by vanishing pressure dependence of the wavefront speed.
\section{Scattering}
\label{sec:scattering}
We now focus on wave scattering and deviations from EMT. First of all we have to distinguish between the coherent and incoherent part of the measured signal. For this purpose we measured signals in 64 packings, each prepared using the same protocol. Before each measurement, the confinement pressure, defined as the average pressure on one sidewall, was relaxed to 500 Pa, then increased to 1 kPa within an error of 10 Pa. The packing was stabilized as described in section \emph{Packing Preparation}. Then a series of Gaussian tone bursts with center frequencies from 10 to 30 kHz, 4 cycles each, was transmitted. The resulting signal after propagation through the packing was measured by the two accelerometers within the packing and two piezos at the wall opposite to the voice-coil. Each measured waveform contains the entire burst series. A high-pass filter of 2 kHz and a low-pass filter of 50 kHz was applied to remove noise outside the relevant bandwidth of this measurement. Then the ensemble-averaging was performed, yielding the coherent part of the signal.

\begin{figure}
\includegraphics[width=\hsize]{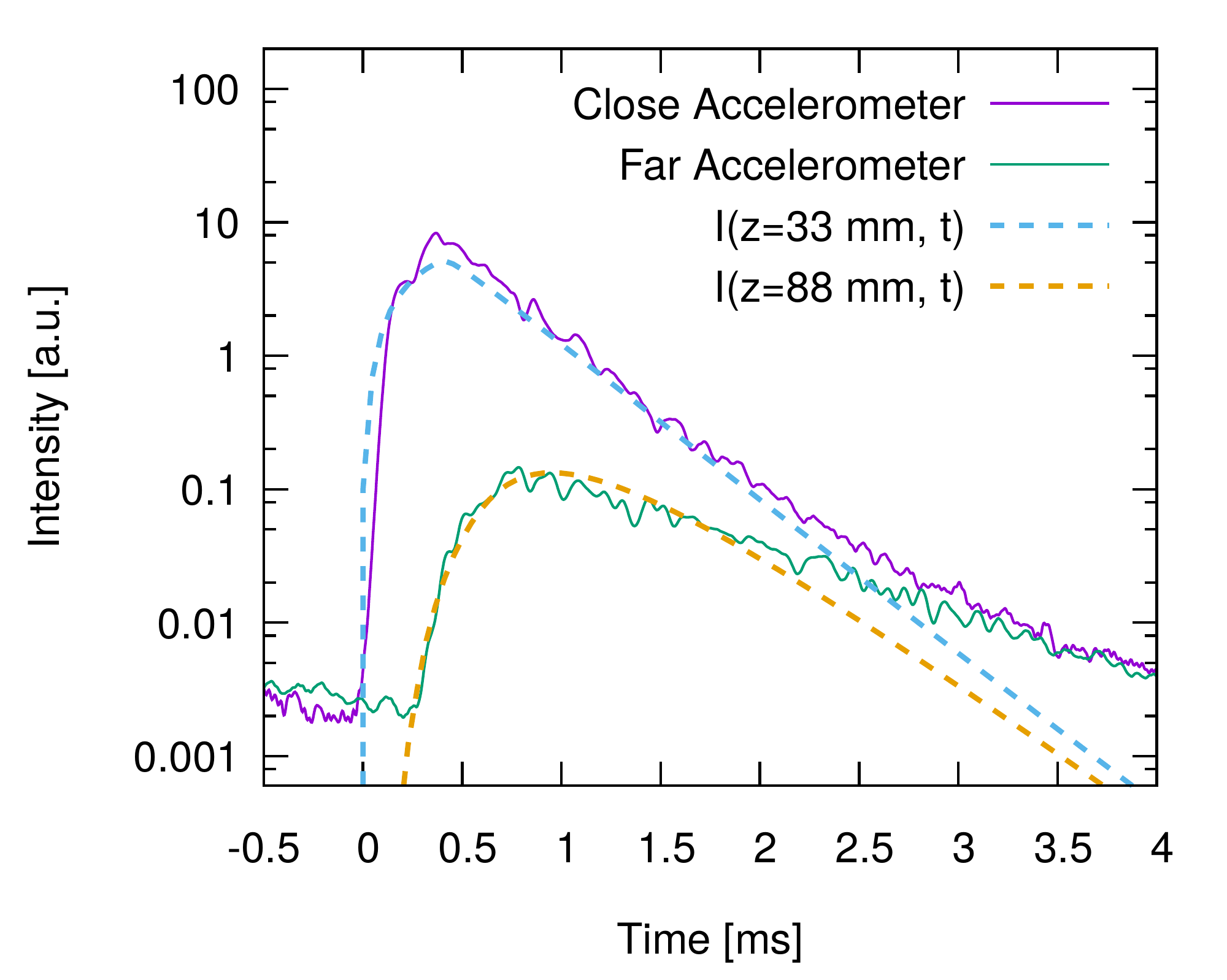}
\caption{Ensemble-averaged intensity of the incoherent received signal (solid lines) after excitation using a 4 cycle Gaussian tone burst of 12 kHz center frequency. The dashed lines show the fitted intensity profile according to Eq. (\ref{eqn:coda_intensity}) at sensor position z after convolution with the tone-burst envelope.}
\label{fig:intensity_fit}
\end{figure}

The incoherent part is obtained by subtracting the coherent part from each waveform. We calculate the intensity of the incoherent part using $I(t)=\|s(t)\|^2$ where $s(t)$ is the analytical signal given by Eq. (\ref{eqn:analytic_signal1})

\begin{align}
s(t)=x(t)+i\cdot H\left[x(t)\right]
\label{eqn:analytic_signal1}
\end{align}

where we use the Hilbert transform:

\begin{align}
H\left[f(t)\right]=\frac{1}{\pi}p.v.\int_{-\infty}^{\infty}\frac{f(\xi)}{t-\xi}d\xi
\label{eqn:hilbert}
\end{align}

and we average the results over all waveforms. The result is plotted in Fig. \ref{fig:intensity_fit} for one particular burst frequency. Here we also show a fit to the predicted intensity of multiply scattered waves where we used a diffusion model\cite{Jia2} taking into account the distance $z$ of each sensor from the vibrating wall, given by:

\begin{align}
I(z,t)=\frac{\nu_e U_0}{2 L}e^{-t / \tau_a}\sum_{n=0}^\infty \frac{1}{\delta_n} cos\left(\frac{n \pi z}{L}\right) cos\left(\frac{n \pi l^\ast }{L}\right) e^{-t D \left(n \pi / L \right)^2}
\label{eqn:coda_intensity}
\end{align}

Before performing the fit, we convoluted the intensity given by Eq. (\ref{eqn:coda_intensity}) with the Gaussian envelope of the excitation signal. Here we identified $\nu_e$, the speed of energy transport, with the sound speed for shear waves, which we approximated as $c_s \approx c_p / \sqrt{3} \approx$ 144 m/s. $\tau_a$ is the inelastic absorption time, which is fitted to 0.4 ms or, equivalently, a quality factor $Q = 2\pi f \tau_a =$ 30. $L$ is the total length of the sample cell, $l^\ast$ is the transport mean free path that we fitted to $l^\ast = 5\cdot d$. $\delta_n$ is $2$ for $n=0$ and $1$ otherwise. $U_0$ is a fit parameter corresponding to the maximum intensity and $D=\nu_e \cdot l^\ast / 3$ is the diffusion coefficient that we fitted to 0.77 $m^2/s$. We have to stress again that our measurements on ground are affected by inhomogeneities in the packing due to gravity and can thus not immediately be compared to previous work\cite{Jia2} where the author applied sufficient confinement pressure that the hydrostatic gradient became negligible.

The ground-based measurement of 64 of newly prepared packings, as shown here, takes about one hour with our apparatus which is well beyond the limits of available time in microgravity in a Mapheus sounding rocket flight. To acquire data of sufficiently large ensembles in microgravity one has to consider adapting the experiment for orbital platforms such as the International Space Station (ISS).

\begin{figure}
\includegraphics[width=\hsize]{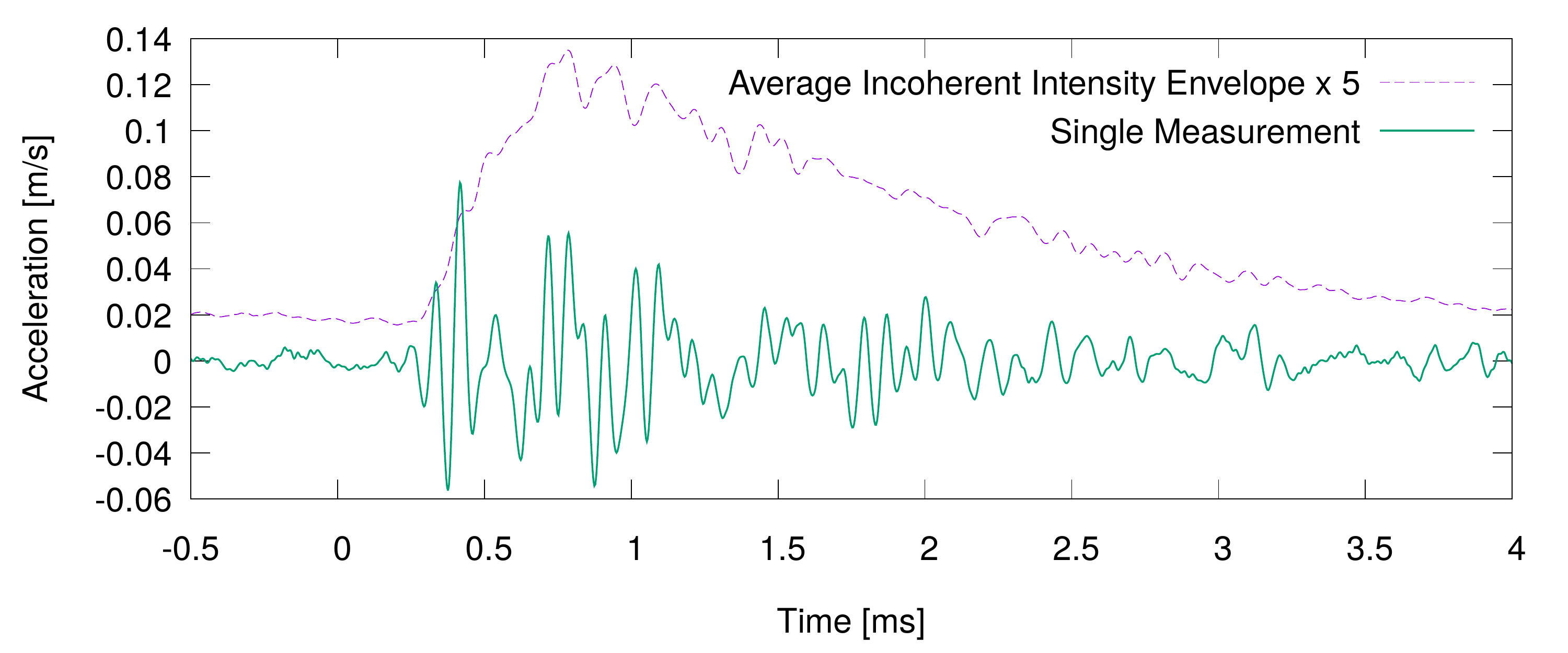}
\caption{Accelerometer signal after excitation using a 4 cycle Gaussian tone burst of 12 kHz center frequency. Green line: signal as received for one of 64 packing configurations. Dashed line: square root of configurationally averaged intensity after subtraction of the average signal from each configuration specific signal.}
\label{fig:raw_and_intensity}
\end{figure}

\section{Conclusion}
\label{sec:conc}
Measurements of acoustic waves in granular packings under low, confinement pressure without hydrostatic gradient require experiments in microgravity. We have developed an apparatus for fully automated granular sound measurements as payload for the Mapheus sounding rocket of DLR. We have demonstrated packing preparation at specific pressure which we tested on ground and in previous microgravity campaigns. Furthermore we have demonstrated a voice coil based excitation system capable of generating both strong, short pulses that lead to shock-wave like behavior as well as sinusoidal or tone burst signals in a large frequency range to probe the transmission of linear waves.

\begin{acknowledgments}
We would like to thank master students Alex Kamphuis and Antoine Micallef for their work on previous granular sound experiments on parabolic flights and on ground. We thank ZARM and NOVESPACE teams for providing drop tower and parabolic flight facilities for our experiments. We appreciate the financial and administrative support from DLR, under project Nos. 50WM1761 and 50WM1651 for the construction of the experimental setups and the usage of the sounding rocket.
\end{acknowledgments}

\appendix
\numberwithin{equation}{section}
\section{Pulse Shaping}
\label{sec:pulse_shaping}
Our excitation system, consisting of signal generator, power amplifier, voice coil and a thin, rigid plate that is driven to vibrate against the granular packing, has a nontrivial transfer function. Each of the mentioned components has a finite bandwidth, a set of resonant modes and frequency dependant attenuation and phase shift. For example, the oscillating plate has resonances determined by its geometric dimensions and its material's speed of sound. Furthermore, the voice coil acts as low-pass filter in two different ways: First of all, mechanically, with a characteristic frequency determined by its spring constant and mass of the moving coil. Second of all, electronically, with a characteristic frequency determined by its inductance and resistance. At high amplitudes, there are also nonlinear distortions. For example, the amplifier is based on semiconductor components such as transistors. Emitter-basis-diodes have exponential current-voltage functions. If correctly biased, the function is approximately linear for sufficiently small perturbations from the bias voltage. For increasingly higher input AC voltage the output contains larger contributions from higher-order terms. For sound amplification this results in higher harmonics. Moreover, extremely large input signals drive the amplifier into saturation, resulting in clipping, which creates further higher harmonics. Analogous behavior occurs for any real driven oscillator. All these signal-altering effects can be reduced to acceptable levels by carefully selecting all components to match the intended application, but the signal range in terms of amplitude and bandwidth is always limited.

To increase the experimentally accessible range of possible excitation signals, we use the following approach:

Instead of fine-tuning each individual component for a given signal type, we treat the entire excitation system at once. At lowest order, the entire chain from the signal generator to a sensor at the vibrating wall can be treated as a damped linear driven oscillator. It is defined by its impulse response function $h(t)$ in the time domain or the corresponding frequency response (and phase response) function in the Fourier domain. For any given input signal $x(t)$ the output $y(t)$ is given by the convolution:

\begin{equation}\begin{aligned}
y(t)=(x\ast h) (t) = \int_{-\infty}^{t}x(t')h(t-t') dt'
\end{aligned}
\label{eqn:lin_resp_conv}
\end{equation}

We make use of the convolution theorem:

\begin{equation}\begin{aligned}
(x\ast h)(t) = \frac{1} {2 \pi} \int_{-\infty}^{\infty}\hat{x}(\omega)\cdot \hat{h}(\omega) e^{i\omega t}d\omega
\end{aligned}
\label{eqn:conv}
\end{equation}

Instead of feeding the desired signal $x(t)$ to the excitation system, we use the filtered signal $x'(t)$, which is obtained by the convolution:

\begin{equation}\begin{aligned}
x'(t)=(x\ast g) (t) = \int_{-\infty}^t x(t')g(t-t') dt'
\end{aligned}
\label{eqn:x_conv_g}
\end{equation}

where $g(t)$ satisfies:

\begin{equation}\begin{aligned}
(x'\ast h)(t) = ((x \ast g)\ast h)(t) = x(t)
\end{aligned}
\label{eqn:x_conv_g_conv_h}
\end{equation}

If $h(t)$ is known, $g(t)$ can be obtained numerically.

To obtain $h(t)$ empirically, we transmit a probe signal $x(t)$ at low amplitude and measure the system output $y(t)$ with an accelerometer mounted on the vibrating wall. The bandwidth of $x(t)$ is chosen to be higher than the excitation system's bandwidth, or, at least as high as any possible excitation signal's bandwidth, including higher harmonics that might affect granular sound measurements. To determine $h(t)$ from this data, a widely used approach is to directly fit a function with sufficiently many parameters in the time-domain. The form of this function is determined by a-priori knowledge of the system's properties. For example, for a system with evenly spaced resonances $\omega_k$ and for $x(t)=\delta(t)$ the expansion

\begin{equation}\begin{aligned}
y(t)=h(t)=\sum_{k=1}^N sin(t \omega_0 k)\cdot\Theta(t)\cdot e^{-\frac{t}{\tau_k}}
\end{aligned}
\label{eqn:h_time}
\end{equation}

can be used with a fundamental resonance $\omega_0$ and a cutoff frequency at $N$ determined by the measurable bandwidth. In more realistic cases, where unknown unrelated resonances of different origin are present, more general functions such as polynomials can be fitted.

Instead, we use an approach in the Fourier domain, that is more easily applicable to any $x(t)$ of arbitrary signal shape and requires no a-priori knowledge of the system other than its bandwidth. To numerically obtain the Fourier transform of measured signals we implemented a variant of the Cooley-Tukey algorithm in our custom signal processing and data analysis program. In the Fourier domain, the empirical transfer function estimate (ETFE) is given by:

\begin{equation}\begin{aligned}
\hat{h}(\omega)=\frac{\hat{y}(\omega)}{\hat{x}(\omega)}
\end{aligned}
\label{eqn:naive_fourier_h}
\end{equation}

However, Eq. (\ref{eqn:naive_fourier_h}) diverges at any points where $\hat{x}(\omega_i)=0$. Such points have to be excluded from the calculation. But even when $\hat{x}(\omega)$ is finite but very small at isolated points $\omega_i$ then $\hat{h}(\omega)$ will be dominated by contributions of finite noise in the output signal $\hat{y}(\omega)$. Furthermore, the measured output in an extended frequency range beyond the excitation bandwidth is dominated by noise. To address these two issues, we use the following techniques:

First of all, we note that the ETFE given by Eq. (\ref{eqn:naive_fourier_h}) is equivalent to:

\begin{equation}\begin{aligned}
\hat{h}(\omega)=\frac{\hat{y}(\omega)\hat{x}^\ast(\omega)}{\hat{x}(\omega)\hat{x}^\ast(\omega)}=\frac{R_{y x}(\omega)}{R_x(\omega)}
\end{aligned}
\label{eqn:fourier_corr_h}
\end{equation}

where the nominator is the cross-spectrum of input and output signals while the denominator is the power spectral density of the input signal. To smear out any isolated divergencies we use a frequency-smoothening technique with a smooth integration kernel $W(\omega)$ that falls off sufficiently fast:

\begin{equation}\begin{aligned}
\hat{h}_{smooth}(\omega)=\frac{\int_{-b}^b R_{y x} (\omega-\omega')\cdot W(\omega') d\omega'}{\int_{-b}^b R_{x} (\omega-\omega')\cdot W(\omega') d\omega'}\\
=\frac{(R_{y x} \ast W)(\omega)}{(R_x \ast W)(\omega)}
\end{aligned}\label{eqn:fourier_smooth_h}
\end{equation}

where $[-b; b]$ is chosen according to the width of $W(\omega)$. The latter is implemented as a Gaussian of a width chosen such that a number of $N_b$ frequency bins are included. Now we assume the output noise is uncorrelated with the input signal. This is justified if $\hat{x}(\omega)$ is zero at $\omega_i$ and small within $[\omega_i -b;\omega_i +b]$ because here $\hat{y}(\omega)$ should contain only (white) noise originating e.g. from any electronic component in the measurement chain. Then this results in a noise reduction of $\propto 1/\sqrt{N_b}$ and effective removal of isolated singularities of the ETFE. It is worth noting that this frequency-smoothening is equivalent to multiplication by a window of width $1/b$ which effectively restricts $h(t)$ to an interval close to $t=0$ and removes unphysical contributions for much later times and anti-causal contributions for $t<<-1/b$.

Secondly, to address the noise contribution at frequencies where $\hat{x}(\omega)$ is nonzero, we introduce a term for the input noise:

\begin{equation}\begin{aligned}
\hat{h}_{smooth, noise}(\omega)=\frac{(R_{y x} \ast W)(\omega)}{N_x+(R_x \ast W)(\omega)}
\end{aligned}
\label{eqn:fourier_smooth_noise_h}
\end{equation}

where $N_x$ is the power spectral density of the (white) input noise. This keeps contributions to the ETFE finite outside the bandwidth of the input signal. Additionally we apply a low-pass filter of 50 kHz to remove frequencies far beyond our system bandwidth.

Once a good linear response estimator is found, as we can verify by forward convolution of $x(t)$ and comparison with $y(t)$, we can obtain $x'(t)$ by

\begin{equation}\begin{aligned}
\hat{x'}(\omega)=\frac{\hat{x}(\omega)}{\hat{h}(\omega)}
\end{aligned}
\label{eqn:inv_fil_fourier}
\end{equation}

while making use of Eq. (\ref{eqn:conv}) and applying an inverse FFT.

Different probe signals are used. To cover the entire system bandwidth, either a sufficiently short pulse, finite duration white noise, a pseudorandom binary sequence (PRBS) or a frequency sweep (chirp) signal can be used. A particularly suitable signal with constant power spectral density and constant frequency resolution per frequency is the white exponential chirp:

\begin{equation}\begin{aligned}
x(t)=\left(\frac{f_{max}}{f_{min}}\right)^{t/(2 T)} \cdot sin\left(2 \pi T \frac{f_{min}\left(\frac{f_{max}}{f_{min}}\right)^{t/T}}{log\left(\frac{f_{max}}{f_{min}}\right)}\right)
\end{aligned}
\label{eqn:exp_chirp}
\end{equation}

where the frequency increases from $f_{min}$ to $f_{max}$ within the signal duration $T$. The prefactor ensures the signal is white.

We can also use the desired output signal, such as a series of short Gaussian pulses, directly as probe signal.

\begin{figure}
\resizebox{\hsize}{!}{\includegraphics{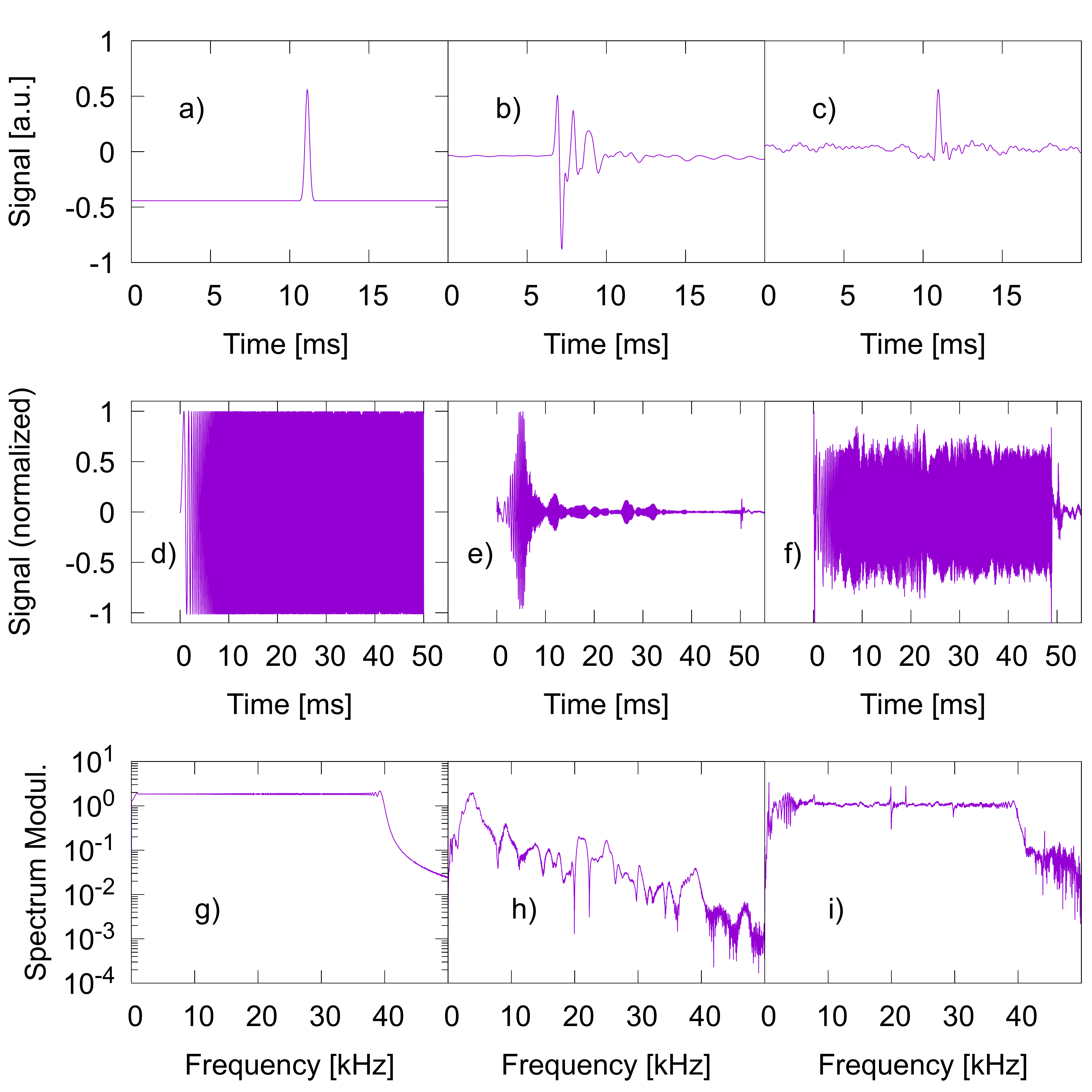}}

\caption{Desired (a,d) and measured excitation signals without (b,e) and with (c,f) inverse filtering before applying the waveform to the signal generator. The narrow pulse (a) is distorted by the excitation system (b). A much cleaner pulse (c) is obtained using inverse filtering by the system's transfer function. A similar improvement is shown for a linear chirp (d-f). The flat spectrum of this signal (g) is lost due to the system's resonances and its low-pass filter behavior (h) but can be almost recovered using the inverse filtering method (i).}
\label{fig:inv_fil_tests}
\end{figure}

In figure \ref{fig:inv_fil_tests} we show measurements of narrow pulse and exponential chirp signals with and without any applied inverse filtering. It can be seen, that despite the narrow bandwidth of the voice coil of only 8 kHz and strong, narrow resonances, we can generate excitation signals having a nearly flat spectrum up to 40 kHz if so desired. For the exponential chirp, it is shown that the signal phase is correct for all frequencies. In this bandwidth, we can now generate arbitrary signal shapes. It must be noted, that this increased effective bandwidth comes at the cost of much lower amplitude range.

%


\begin{thebibliography}{24}%
\makeatletter
\providecommand \@ifxundefined [1]{%
 \@ifx{#1\undefined}
}%
\providecommand \@ifnum [1]{%
 \ifnum #1\expandafter \@firstoftwo
 \else \expandafter \@secondoftwo
 \fi
}%
\providecommand \@ifx [1]{%
 \ifx #1\expandafter \@firstoftwo
 \else \expandafter \@secondoftwo
 \fi
}%
\providecommand \natexlab [1]{#1}%
\providecommand \enquote  [1]{``#1''}%
\providecommand \bibnamefont  [1]{#1}%
\providecommand \bibfnamefont [1]{#1}%
\providecommand \citenamefont [1]{#1}%
\providecommand \href@noop [0]{\@secondoftwo}%
\providecommand \href [0]{\begingroup \@sanitize@url \@href}%
\providecommand \@href[1]{\@@startlink{#1}\@@href}%
\providecommand \@@href[1]{\endgroup#1\@@endlink}%
\providecommand \@sanitize@url [0]{\catcode `\\12\catcode `\$12\catcode
  `\&12\catcode `\#12\catcode `\^12\catcode `\_12\catcode `\%12\relax}%
\providecommand \@@startlink[1]{}%
\providecommand \@@endlink[0]{}%
\providecommand \url  [0]{\begingroup\@sanitize@url \@url }%
\providecommand \@url [1]{\endgroup\@href {#1}{\urlprefix }}%
\providecommand \urlprefix  [0]{URL }%
\providecommand \Eprint [0]{\href }%
\providecommand \doibase [0]{https://doi.org/}%
\providecommand \selectlanguage [0]{\@gobble}%
\providecommand \bibinfo  [0]{\@secondoftwo}%
\providecommand \bibfield  [0]{\@secondoftwo}%
\providecommand \translation [1]{[#1]}%
\providecommand \BibitemOpen [0]{}%
\providecommand \bibitemStop [0]{}%
\providecommand \bibitemNoStop [0]{.\EOS\space}%
\providecommand \EOS [0]{\spacefactor3000\relax}%
\providecommand \BibitemShut  [1]{\csname bibitem#1\endcsname}%
\let\auto@bib@innerbib\@empty
\bibitem [{\citenamefont {Liu}\ and\ \citenamefont {Nagel}(1993)}]{Liu_Nagel}%
  \BibitemOpen
  \bibfield  {author} {\bibinfo {author} {\bibfnamefont {C.-h.}\ \bibnamefont
  {Liu}}\ and\ \bibinfo {author} {\bibfnamefont {S.~R.}\ \bibnamefont
  {Nagel}},\ }\bibfield  {title} {\enquote {\bibinfo {title} {Sound in a
  granular material: Disorder and nonlinearity},}\ }\href
  {https://doi.org/10.1103/PhysRevB.48.15646} {\bibfield  {journal} {\bibinfo
  {journal} {Phys. Rev. B}\ }\textbf {\bibinfo {volume} {48}},\ \bibinfo
  {pages} {15646--15650} (\bibinfo {year} {1993})}\BibitemShut {NoStop}%
\bibitem [{\citenamefont {Jia}, \citenamefont {Caroli},\ and\ \citenamefont
  {Velicky}(1999)}]{Jia1}%
  \BibitemOpen
  \bibfield  {author} {\bibinfo {author} {\bibfnamefont {X.}~\bibnamefont
  {Jia}}, \bibinfo {author} {\bibfnamefont {C.}~\bibnamefont {Caroli}},\ and\
  \bibinfo {author} {\bibfnamefont {B.}~\bibnamefont {Velicky}},\ }\bibfield
  {title} {\enquote {\bibinfo {title} {Ultrasound propagation in externally
  stressed granular media},}\ }\href
  {https://doi.org/10.1103/PhysRevLett.82.1863} {\bibfield  {journal} {\bibinfo
   {journal} {Phys. Rev. Lett.}\ }\textbf {\bibinfo {volume} {82}},\ \bibinfo
  {pages} {1863--1866} (\bibinfo {year} {1999})}\BibitemShut {NoStop}%
\bibitem [{\citenamefont {Walton}(1987)}]{Walton}%
  \BibitemOpen
  \bibfield  {author} {\bibinfo {author} {\bibfnamefont {K.}~\bibnamefont
  {Walton}},\ }\bibfield  {title} {\enquote {\bibinfo {title} {The effective
  elastic moduli of a random packing of spheres},}\ }\href
  {https://doi.org/https://doi.org/10.1016/0022-5096(87)90036-6} {\bibfield
  {journal} {\bibinfo  {journal} {J. Mech. Phys. Solids}\ }\textbf {\bibinfo
  {volume} {35}},\ \bibinfo {pages} {213 -- 226} (\bibinfo {year}
  {1987})}\BibitemShut {NoStop}%
\bibitem [{\citenamefont {Digby}(1981)}]{Digby}%
  \BibitemOpen
  \bibfield  {author} {\bibinfo {author} {\bibfnamefont {P.~J.}\ \bibnamefont
  {Digby}},\ }\bibfield  {title} {\enquote {\bibinfo {title} {The effective
  elastic moduli of porous granular rocks},}\ }\href
  {https://doi.org/10.1115/1.3157738} {\bibfield  {journal} {\bibinfo
  {journal} {J. Appl. Mech.}\ }\textbf {\bibinfo {volume} {48}},\ \bibinfo
  {pages} {803} (\bibinfo {year} {1981})}\BibitemShut {NoStop}%
\bibitem [{\citenamefont {Goddard}(1990)}]{Goddard}%
  \BibitemOpen
  \bibfield  {author} {\bibinfo {author} {\bibfnamefont {J.}~\bibnamefont
  {Goddard}},\ }\bibfield  {title} {\enquote {\bibinfo {title} {Nonlinear
  elasticity and pressure-dependent wave speeds in granular media},}\ }\href
  {https://doi.org/10.1098/rspa.1990.0083} {\bibfield  {journal} {\bibinfo
  {journal} {Proceedings of The Royal Society A: Mathematical, Physical and
  Engineering Sciences}\ }\textbf {\bibinfo {volume} {430}},\ \bibinfo {pages}
  {105--131} (\bibinfo {year} {1990})}\BibitemShut {NoStop}%
\bibitem [{\citenamefont {Coste}\ and\ \citenamefont {Gilles}(2008)}]{Coste}%
  \BibitemOpen
  \bibfield  {author} {\bibinfo {author} {\bibfnamefont {C.}~\bibnamefont
  {Coste}}\ and\ \bibinfo {author} {\bibfnamefont {B.}~\bibnamefont {Gilles}},\
  }\bibfield  {title} {\enquote {\bibinfo {title} {Sound propagation in a
  constrained lattice of beads: High-frequency behavior and dispersion
  relation},}\ }\href {https://doi.org/10.1103/PhysRevE.77.021302} {\bibfield
  {journal} {\bibinfo  {journal} {Phys. Rev. E}\ }\textbf {\bibinfo {volume}
  {77}},\ \bibinfo {pages} {021302} (\bibinfo {year} {2008})}\BibitemShut
  {NoStop}%
\bibitem [{\citenamefont {Jia}(2004)}]{Jia2}%
  \BibitemOpen
  \bibfield  {author} {\bibinfo {author} {\bibfnamefont {X.}~\bibnamefont
  {Jia}},\ }\bibfield  {title} {\enquote {\bibinfo {title} {Codalike multiple
  scattering of elastic waves in dense granular media},}\ }\href
  {https://doi.org/10.1103/PhysRevLett.93.154303} {\bibfield  {journal}
  {\bibinfo  {journal} {Phys. Rev. Lett.}\ }\textbf {\bibinfo {volume} {93}},\
  \bibinfo {pages} {154303} (\bibinfo {year} {2004})}\BibitemShut {NoStop}%
\bibitem [{\citenamefont {Travers}\ \emph {et~al.}(1987)\citenamefont
  {Travers}, \citenamefont {Ammi}, \citenamefont {Bideau}, \citenamefont
  {Gervois}, \citenamefont {Messager},\ and\ \citenamefont
  {Troadec}}]{travers}%
  \BibitemOpen
  \bibfield  {author} {\bibinfo {author} {\bibfnamefont {T.}~\bibnamefont
  {Travers}}, \bibinfo {author} {\bibfnamefont {M.}~\bibnamefont {Ammi}},
  \bibinfo {author} {\bibfnamefont {D.}~\bibnamefont {Bideau}}, \bibinfo
  {author} {\bibfnamefont {A.}~\bibnamefont {Gervois}}, \bibinfo {author}
  {\bibfnamefont {J.}~\bibnamefont {Messager}},\ and\ \bibinfo {author}
  {\bibfnamefont {J.}~\bibnamefont {Troadec}},\ }\bibfield  {title} {\enquote
  {\bibinfo {title} {Uniaxial compression of 2d packings of cylinders. effects
  of weak disorder},}\ }\href@noop {} {\bibfield  {journal} {\bibinfo
  {journal} {EPL}\ }\textbf {\bibinfo {volume} {4}},\ \bibinfo {pages} {329}
  (\bibinfo {year} {1987})}\BibitemShut {NoStop}%
\bibitem [{\citenamefont {Radjai}\ \emph {et~al.}(1996)\citenamefont {Radjai},
  \citenamefont {Jean}, \citenamefont {Moreau},\ and\ \citenamefont
  {Roux}}]{radjai}%
  \BibitemOpen
  \bibfield  {author} {\bibinfo {author} {\bibfnamefont {F.}~\bibnamefont
  {Radjai}}, \bibinfo {author} {\bibfnamefont {M.}~\bibnamefont {Jean}},
  \bibinfo {author} {\bibfnamefont {J.-J.}\ \bibnamefont {Moreau}},\ and\
  \bibinfo {author} {\bibfnamefont {S.}~\bibnamefont {Roux}},\ }\bibfield
  {title} {\enquote {\bibinfo {title} {Force distributions in dense
  two-dimensional granular systems},}\ }\href
  {https://doi.org/10.1103/PhysRevLett.77.274} {\bibfield  {journal} {\bibinfo
  {journal} {Phys. Rev. Lett.}\ }\textbf {\bibinfo {volume} {77}},\ \bibinfo
  {pages} {274--277} (\bibinfo {year} {1996})}\BibitemShut {NoStop}%
\bibitem [{\citenamefont {Jaeger}, \citenamefont {Nagel},\ and\ \citenamefont
  {Behringer}(1996)}]{jaeger}%
  \BibitemOpen
  \bibfield  {author} {\bibinfo {author} {\bibfnamefont {H.~M.}\ \bibnamefont
  {Jaeger}}, \bibinfo {author} {\bibfnamefont {S.~R.}\ \bibnamefont {Nagel}},\
  and\ \bibinfo {author} {\bibfnamefont {R.~P.}\ \bibnamefont {Behringer}},\
  }\bibfield  {title} {\enquote {\bibinfo {title} {Granular solids, liquids,
  and gases},}\ }\href@noop {} {\bibfield  {journal} {\bibinfo  {journal}
  {Reviews of modern physics}\ }\textbf {\bibinfo {volume} {68}},\ \bibinfo
  {pages} {1259} (\bibinfo {year} {1996})}\BibitemShut {NoStop}%
\bibitem [{\citenamefont {Owens}\ and\ \citenamefont
  {Daniels}(2011)}]{Daniels1}%
  \BibitemOpen
  \bibfield  {author} {\bibinfo {author} {\bibfnamefont {E.~T.}\ \bibnamefont
  {Owens}}\ and\ \bibinfo {author} {\bibfnamefont {K.~E.}\ \bibnamefont
  {Daniels}},\ }\bibfield  {title} {\enquote {\bibinfo {title} {Sound
  propagation and force chains in granular materials},}\ }\href
  {https://doi.org/10.1209/0295-5075/94/54005} {\bibfield  {journal} {\bibinfo
  {journal} {EPL}\ }\textbf {\bibinfo {volume} {94}},\ \bibinfo {pages} {54005}
  (\bibinfo {year} {2011})}\BibitemShut {NoStop}%
\bibitem [{\citenamefont {Tournat}\ and\ \citenamefont
  {Gusev}(2009)}]{Tournat1}%
  \BibitemOpen
  \bibfield  {author} {\bibinfo {author} {\bibfnamefont {V.}~\bibnamefont
  {Tournat}}\ and\ \bibinfo {author} {\bibfnamefont {V.~E.}\ \bibnamefont
  {Gusev}},\ }\bibfield  {title} {\enquote {\bibinfo {title} {Nonlinear effects
  for coda-type elastic waves in stressed granular media},}\ }\href
  {https://doi.org/10.1103/PhysRevE.80.011306} {\bibfield  {journal} {\bibinfo
  {journal} {Phys. Rev. E}\ }\textbf {\bibinfo {volume} {80}},\ \bibinfo
  {pages} {011306} (\bibinfo {year} {2009})}\BibitemShut {NoStop}%
\bibitem [{\citenamefont {Jia}, \citenamefont {Brunet},\ and\ \citenamefont
  {Laurent}(2011)}]{Jia3}%
  \BibitemOpen
  \bibfield  {author} {\bibinfo {author} {\bibfnamefont {X.}~\bibnamefont
  {Jia}}, \bibinfo {author} {\bibfnamefont {T.}~\bibnamefont {Brunet}},\ and\
  \bibinfo {author} {\bibfnamefont {J.}~\bibnamefont {Laurent}},\ }\bibfield
  {title} {\enquote {\bibinfo {title} {Elastic weakening of a dense granular
  pack by acoustic fluidization: Slipping, compaction, and aging},}\ }\href
  {https://doi.org/10.1103/PhysRevE.84.020301} {\bibfield  {journal} {\bibinfo
  {journal} {Phys. Rev. E}\ }\textbf {\bibinfo {volume} {84}},\ \bibinfo
  {pages} {020301} (\bibinfo {year} {2011})}\BibitemShut {NoStop}%
\bibitem [{\citenamefont {Trujillo}, \citenamefont {Peniche},\ and\
  \citenamefont {Jia}(2011)}]{Trujillo}%
  \BibitemOpen
  \bibfield  {author} {\bibinfo {author} {\bibfnamefont {L.}~\bibnamefont
  {Trujillo}}, \bibinfo {author} {\bibfnamefont {F.}~\bibnamefont {Peniche}},\
  and\ \bibinfo {author} {\bibfnamefont {X.}~\bibnamefont {Jia}},\ }\bibfield
  {title} {\enquote {\bibinfo {title} {Multiple scattering of elastic waves in
  granular media: Theory and experiments},}\ }in\ \href
  {https://doi.org/10.5772/17707} {\emph {\bibinfo {booktitle} {Waves in Fluids
  and Solids}}},\ \bibinfo {editor} {edited by\ \bibinfo {editor}
  {\bibfnamefont {R.~P.}\ \bibnamefont {Vila}}}\ (\bibinfo  {publisher}
  {IntechOpen},\ \bibinfo {address} {Rijeka},\ \bibinfo {year} {2011})\
  Chap.~\bibinfo {chapter} {5}\BibitemShut {NoStop}%
\bibitem [{\citenamefont {Nesterenko}(1983)}]{Nesterenko}%
  \BibitemOpen
  \bibfield  {author} {\bibinfo {author} {\bibfnamefont {V.}~\bibnamefont
  {Nesterenko}},\ }\bibfield  {title} {\enquote {\bibinfo {title} {Propagation
  of nonlinear compression pulses in granular media},}\ }\href@noop {}
  {\bibfield  {journal} {\bibinfo  {journal} {J. Appl. Mech. Tech. Phys. (Engl.
  Transl.)}\ }\textbf {\bibinfo {volume} {24}},\ \bibinfo {pages} {733--743}
  (\bibinfo {year} {1983})}\BibitemShut {NoStop}%
\bibitem [{\citenamefont {Daraio}\ \emph {et~al.}(2005)\citenamefont {Daraio},
  \citenamefont {Nesterenko}, \citenamefont {Herbold},\ and\ \citenamefont
  {Jin}}]{Daraio_1dchain}%
  \BibitemOpen
  \bibfield  {author} {\bibinfo {author} {\bibfnamefont {C.}~\bibnamefont
  {Daraio}}, \bibinfo {author} {\bibfnamefont {V.~F.}\ \bibnamefont
  {Nesterenko}}, \bibinfo {author} {\bibfnamefont {E.~B.}\ \bibnamefont
  {Herbold}},\ and\ \bibinfo {author} {\bibfnamefont {S.}~\bibnamefont {Jin}},\
  }\bibfield  {title} {\enquote {\bibinfo {title} {Strongly nonlinear waves in
  a chain of teflon beads},}\ }\href
  {https://doi.org/10.1103/PhysRevE.72.016603} {\bibfield  {journal} {\bibinfo
  {journal} {Phys. Rev. E}\ }\textbf {\bibinfo {volume} {72}},\ \bibinfo
  {pages} {016603} (\bibinfo {year} {2005})}\BibitemShut {NoStop}%
\bibitem [{\citenamefont {G\'omez}\ \emph {et~al.}(2012)\citenamefont
  {G\'omez}, \citenamefont {Turner}, \citenamefont {van Hecke},\ and\
  \citenamefont {Vitelli}}]{Gomez}%
  \BibitemOpen
  \bibfield  {author} {\bibinfo {author} {\bibfnamefont {L.~R.}\ \bibnamefont
  {G\'omez}}, \bibinfo {author} {\bibfnamefont {A.~M.}\ \bibnamefont {Turner}},
  \bibinfo {author} {\bibfnamefont {M.}~\bibnamefont {van Hecke}},\ and\
  \bibinfo {author} {\bibfnamefont {V.}~\bibnamefont {Vitelli}},\ }\bibfield
  {title} {\enquote {\bibinfo {title} {Shocks near jamming},}\ }\href
  {https://doi.org/10.1103/PhysRevLett.108.058001} {\bibfield  {journal}
  {\bibinfo  {journal} {Phys. Rev. Lett.}\ }\textbf {\bibinfo {volume} {108}},\
  \bibinfo {pages} {058001} (\bibinfo {year} {2012})}\BibitemShut {NoStop}%
\bibitem [{\citenamefont {van~den Wildenberg}, \citenamefont {van Loo},\ and\
  \citenamefont {van Hecke}(2013)}]{Wildenberg}%
  \BibitemOpen
  \bibfield  {author} {\bibinfo {author} {\bibfnamefont {S.}~\bibnamefont
  {van~den Wildenberg}}, \bibinfo {author} {\bibfnamefont {R.}~\bibnamefont
  {van Loo}},\ and\ \bibinfo {author} {\bibfnamefont {M.}~\bibnamefont {van
  Hecke}},\ }\bibfield  {title} {\enquote {\bibinfo {title} {Shock waves in
  weakly compressed granular media},}\ }\href
  {https://doi.org/10.1103/PhysRevLett.111.218003} {\bibfield  {journal}
  {\bibinfo  {journal} {Phys. Rev. Lett.}\ }\textbf {\bibinfo {volume} {111}},\
  \bibinfo {pages} {218003} (\bibinfo {year} {2013})}\BibitemShut {NoStop}%
\bibitem [{\citenamefont {Santibanez}, \citenamefont {Zu\~niga},\ and\
  \citenamefont {Melo}(2016)}]{Santibanez}%
  \BibitemOpen
  \bibfield  {author} {\bibinfo {author} {\bibfnamefont {F.}~\bibnamefont
  {Santibanez}}, \bibinfo {author} {\bibfnamefont {R.}~\bibnamefont
  {Zu\~niga}},\ and\ \bibinfo {author} {\bibfnamefont {F.}~\bibnamefont
  {Melo}},\ }\bibfield  {title} {\enquote {\bibinfo {title} {Mechanical impulse
  propagation in a three-dimensional packing of spheres confined at constant
  pressure},}\ }\href {https://doi.org/10.1103/PhysRevE.93.012908} {\bibfield
  {journal} {\bibinfo  {journal} {Phys. Rev. E}\ }\textbf {\bibinfo {volume}
  {93}},\ \bibinfo {pages} {012908} (\bibinfo {year} {2016})}\BibitemShut
  {NoStop}%
\bibitem [{\citenamefont {Zeng}, \citenamefont {Agui},\ and\ \citenamefont
  {Nakagawa}(2007)}]{Zeng}%
  \BibitemOpen
  \bibfield  {author} {\bibinfo {author} {\bibfnamefont {X.}~\bibnamefont
  {Zeng}}, \bibinfo {author} {\bibfnamefont {J.~H.}\ \bibnamefont {Agui}},\
  and\ \bibinfo {author} {\bibfnamefont {M.}~\bibnamefont {Nakagawa}},\
  }\bibfield  {title} {\enquote {\bibinfo {title} {Wave velocities in granular
  materials under microgravity},}\ }\href
  {https://doi.org/10.1061/(ASCE)0893-1321(2007)20:2(116)} {\bibfield
  {journal} {\bibinfo  {journal} {J. Aerosp. Eng.}\ }\textbf {\bibinfo {volume}
  {20}},\ \bibinfo {pages} {116--123} (\bibinfo {year} {2007})}\BibitemShut
  {NoStop}%
\bibitem [{\citenamefont {Auma{\^\i}tre}\ \emph {et~al.}(2018)\citenamefont
  {Auma{\^\i}tre}, \citenamefont {Behringer}, \citenamefont {Cazaubiel},
  \citenamefont {Cl{\'e}ment}, \citenamefont {Crassous}, \citenamefont
  {Durian}, \citenamefont {Falcon}, \citenamefont {Fauve}, \citenamefont
  {Fischer}, \citenamefont {Garcimart{\'\i}n} \emph {et~al.}}]{aumaitre2018}%
  \BibitemOpen
  \bibfield  {author} {\bibinfo {author} {\bibfnamefont {S.}~\bibnamefont
  {Auma{\^\i}tre}}, \bibinfo {author} {\bibfnamefont {R.}~\bibnamefont
  {Behringer}}, \bibinfo {author} {\bibfnamefont {A.}~\bibnamefont
  {Cazaubiel}}, \bibinfo {author} {\bibfnamefont {E.}~\bibnamefont
  {Cl{\'e}ment}}, \bibinfo {author} {\bibfnamefont {J.}~\bibnamefont
  {Crassous}}, \bibinfo {author} {\bibfnamefont {D.}~\bibnamefont {Durian}},
  \bibinfo {author} {\bibfnamefont {E.}~\bibnamefont {Falcon}}, \bibinfo
  {author} {\bibfnamefont {S.}~\bibnamefont {Fauve}}, \bibinfo {author}
  {\bibfnamefont {D.}~\bibnamefont {Fischer}}, \bibinfo {author} {\bibfnamefont
  {A.}~\bibnamefont {Garcimart{\'\i}n}}, \emph {et~al.},\ }\bibfield  {title}
  {\enquote {\bibinfo {title} {An instrument for studying granular media in
  low-gravity environment},}\ }\href@noop {} {\bibfield  {journal} {\bibinfo
  {journal} {Review of scientific instruments}\ }\textbf {\bibinfo {volume}
  {89}},\ \bibinfo {pages} {075103} (\bibinfo {year} {2018})}\BibitemShut
  {NoStop}%
\bibitem [{\citenamefont {Siegl}\ \emph {et~al.}(2013)\citenamefont {Siegl},
  \citenamefont {Kargl}, \citenamefont {Scheuerpflug}, \citenamefont
  {Drescher}, \citenamefont {Neumann}, \citenamefont {Balter}, \citenamefont
  {Kolbe}, \citenamefont {Sperl}, \citenamefont {Yu},\ and\ \citenamefont
  {Meyer}}]{mapheus}%
  \BibitemOpen
  \bibfield  {author} {\bibinfo {author} {\bibfnamefont {M.}~\bibnamefont
  {Siegl}}, \bibinfo {author} {\bibfnamefont {F.}~\bibnamefont {Kargl}},
  \bibinfo {author} {\bibfnamefont {F.}~\bibnamefont {Scheuerpflug}}, \bibinfo
  {author} {\bibfnamefont {J.}~\bibnamefont {Drescher}}, \bibinfo {author}
  {\bibfnamefont {C.}~\bibnamefont {Neumann}}, \bibinfo {author} {\bibfnamefont
  {M.}~\bibnamefont {Balter}}, \bibinfo {author} {\bibfnamefont
  {M.}~\bibnamefont {Kolbe}}, \bibinfo {author} {\bibfnamefont
  {M.}~\bibnamefont {Sperl}}, \bibinfo {author} {\bibfnamefont
  {P.}~\bibnamefont {Yu}},\ and\ \bibinfo {author} {\bibfnamefont
  {A.}~\bibnamefont {Meyer}},\ }\bibfield  {title} {\enquote {\bibinfo {title}
  {Material physics rockets mapheus-3/4: flights and developments},}\ }in\
  \href@noop {} {\emph {\bibinfo {booktitle} {Proceedings of the 21st ESA
  Symposium on European Rocket and Balloon Programmes and Related Research}}},\
  Vol.~\bibinfo {volume} {9}\ (\bibinfo {year} {2013})\ p.~\bibinfo {pages}
  {13}\BibitemShut {NoStop}%
\bibitem [{\citenamefont {{van den Wildenberg, S.}}, \citenamefont {{van Hecke,
  M.}},\ and\ \citenamefont {{Jia, X.}}(2013)}]{Jia4}%
  \BibitemOpen
  \bibfield  {author} {\bibinfo {author} {\bibnamefont {{van den Wildenberg,
  S.}}}, \bibinfo {author} {\bibnamefont {{van Hecke, M.}}},\ and\ \bibinfo
  {author} {\bibnamefont {{Jia, X.}}},\ }\bibfield  {title} {\enquote {\bibinfo
  {title} {Evolution of granular packings by nonlinear acoustic waves},}\
  }\href {https://doi.org/10.1209/0295-5075/101/14004} {\bibfield  {journal}
  {\bibinfo  {journal} {EPL}\ }\textbf {\bibinfo {volume} {101}},\ \bibinfo
  {pages} {14004} (\bibinfo {year} {2013})}\BibitemShut {NoStop}%
\bibitem [{\citenamefont {Langlois}\ and\ \citenamefont {Jia}(2015)}]{Jia5}%
  \BibitemOpen
  \bibfield  {author} {\bibinfo {author} {\bibfnamefont {V.}~\bibnamefont
  {Langlois}}\ and\ \bibinfo {author} {\bibfnamefont {X.}~\bibnamefont {Jia}},\
  }\bibfield  {title} {\enquote {\bibinfo {title} {Sound pulse broadening in
  stressed granular media},}\ }\href
  {https://doi.org/10.1103/PhysRevE.91.022205} {\bibfield  {journal} {\bibinfo
  {journal} {Phys. Rev. E}\ }\textbf {\bibinfo {volume} {91}},\ \bibinfo
  {pages} {022205} (\bibinfo {year} {2015})}\BibitemShut {NoStop}%
\end{thebibliography}
\end{document}